\documentclass{PoS}
\usepackage{amsmath,graphicx}
\newcommand{\bra}{\langle}
\newcommand{\ket}{\rangle}
\newcommand{\psibar}{\overline{\psi}}

\newlength{\colw}
\PoS{PoS(LAT2005)176}

\title{Charmonium spectral functions in Nf=2 QCD }

\ShortTitle{Charmonium spectral functions in Nf=2 QCD }

\author{R.~Morrin, A.P.~\'O Cais, M.B.~Oktay, M.J.~Peardon,
 \speaker{J.I.~Skullerud}\\ Trinity College Dublin\\
 E-mail: \email{jonivar@maths.tcd.ie}}

\author{G.~Aarts\thanks{PPARC Advanced Fellow}, C.R.~Allton\\
  University of Wales Swansea}

\abstract{We report on a study of charmonium at high temperature in
  2-flavour QCD.  This is the first such study with dynamical
  fermions.  Using an improved anisotropic lattice action, spectral
  functions are extracted from correlators in the vector and
  pseudoscalar channels.  No signs of medium-induced suppression of
  the ground states are seen for temperatures up to $1.5T_c$, while at
  $T\sim2T_c$ there are clear signs of modifications.  The current
  systematic and statistical uncertainties in our data, in particular
  the relatively coarse lattice and small volume, do not allow us
  to draw a firm conclusion at this stage.}

\FullConference{XXIIIrd International Symposium on Lattice Field Theory\\
		 25-30 July 2005\\
		 Trinity College, Dublin, Ireland}

\begin{document}

\section{Introduction}

The fate of charmonium states in the deconfined phase of QCD has long
been a subject of interest, following the suggestion
\cite{Matsui:1986dk} that $J/\psi$ suppression may be a signature of
deconfinement in heavy-ion collisions.  Potential model calculations
indicated that $J/\psi$ might disappear from the spectrum almost
immediately after deconfinement, providing an unambiguous signature.
Recent lattice calculations
\cite{Umeda:2002vr,Asakawa:2003re,Datta:2003ww} have cast doubt over
this, indicating instead that $J/\psi$ may survive in the plasma up to
temperatures as high as $1.5-2T_c$.  However, all these studies have
been carried out in the quenched approximation, raising serious
questions about their reliability.

The properties of hadrons in the medium are encoded in the spectral
functions $\rho(\omega,\vec{p})$, which are related to the
imaginary-time correlator $G(\tau,\vec{p})$ according to
\begin{equation}
G_\Gamma(\tau,\vec{p}) = \frac{1}{2\pi}
\int_0^\infty\rho_\Gamma(\omega,\vec{p})K(\tau,\omega)d\omega\,,
\label{eq:spectral}
\end{equation}
where
\begin{equation}
K(\tau,\omega) = \frac{\cosh[\omega(\tau-1/2T)]}{\sinh(\omega/2T)}
 = e^{\omega\tau}n_B(\omega) + e^{-\omega\tau}[1+n_B(\omega)]\,.
\label{eq:kernel}
\end{equation}
Here, $n_B$ is the Bose--Einstein distribution function.
Spectral functions may also be used to
extract transport coefficients, while at non-zero momentum
they contain information about phenomena such as Landau damping
\cite{Aarts:2005hg}.

Determining $\rho(\omega)$ from lattice correlators $G(\tau)$ is an
ill-posed problem, but it is possible to get a handle on it using the
Maximum Entropy Method (MEM) \cite{Asakawa:2000tr}.  It is however
crucial to have a sufficient number of points in the euclidean time
direction, and for this purpose anisotropic lattices have a strong
advantage over the more common isotropic formulation.  A further
advantage is that it allows the temperature to be varied in small
steps while keeping the lattice spacing and spatial volume fixed.

\section{Simulation details}

We use the Two-plaquette Symanzik Improved gauge action
\cite{Morningstar:1999dh} and the fine-Wilson, coarse-Hamber-Wu
fermion action \cite{Foley:2004jf} with stout-link smearing
\cite{Morningstar:2003gk}.  The fermion action has been designed with
heavy quarks in mind, and a quenched study \cite{Foley:2004jf} found
that the same bare anisotropy can be used for
valence quark masses ranging from well below strange to well beyond
charm.  All-to-all propagators \cite{Foley:2005ac} with no
eigenvectors and two noise vectors diluted in time, colour and
even/odd in space, were used to
improve the signal.  The parameters correspond to point 5 in
\cite{Morrin:2005xx}.  Using the 1S--1P splitting in charmonium to set
the scale, the lattice spacings were found to be $a_t=0.025$fm,
$a_s=0.2$fm, with the (quark) anisotropy $\xi\sim8$ determined from
the charmonium dispersion relation.  The sea quark mass corresponds to
$m_\pi/m_\rho\approx0.55$.
%table~\ref{tab:param}.

%\begin{table}
%\begin{tabular}{|lll|}
%\hline
%Gauge action & TSI 3+1 & \\
%& $\beta$ & 1.522 \\
%& $\xi_g$ & 7.44 \\
%& $\omega$ & 3.0 \\
%& $u_s^4$ & 0.32 \\\hline
%Fermion action & fWcHW & \\
%& $m_{\text{sea}}$ & -0.057 \\
%& $\xi_q$ & 8.83 \\
%& $u_s, u_t$ & 1 \\
%& Stout-links & 2x0.22 spatial \\
%Charm quarks & $m_c$ & 0.2 \\
%\hline Physical values & & \\
%& $a_s$ & $\sim0.2$ fm \\
%& $\xi_r$ & 6 \\
%& $m_\pi/m_\rho$ & $\sim0.55$ \\\hline
%\end{tabular}
%\caption{Simulation parameters.}
%\label{tab:param}
%\end{table}

Anisotropic lattices with dynamical fermions involve problems that are
not present in the quenched approximation: since the gauge fields
depend on the fermions, it is not possible to first fix the gauge
anisotropy and then tune the fermion anisotropy so that the physical
(measured) anisotropy for the fermions matches that of the gauge
fields.  Instead, the fermion and gauge anisotropies must be tuned
simultaneously \cite{Morrin:2005xx}.  We do not as yet have data at
the fully tuned point given in \cite{Morrin:2005xx}, so unequal quark
and gluon anisotropies are still a significant source of systematic
uncertainty.

We have performed simulations on $N_s^3=8^3$ lattices with $N_t=48,
32, 24$ and 16, corresponding to $T\approx0.75, 1.1, 1.5$ and $2.2T_c$
respectively.  In all three cases we have used 100 configurations,
sampling configurations every 10 HMC trajectories. In addition we have
simulated at $N_t=28,30,33-35$ in an attempt to locate $T_c$.

In this preliminary study we only look at zero momentum.
The euclidean correlators have been computed using local (unsmeared)
operators:
\begin{equation}
G_\Gamma(\tau) =
\sum_{\vec{x},\vec{y},t}\bra\psibar(\vec{x},t)\Gamma\psi(\vec{x},t)
\psibar(\vec{y},t+\tau)\Gamma\psi(\vec{y},t+\tau)\ket \,,
\end{equation}
with $\Gamma=\gamma_5,\gamma_i,1,i\gamma_5\gamma_i$, corresponding to
the pseudoscalar, vector, scalar and axial-vector channel
respectively. The signal in the scalar and axial-vector channels is
poor and those results will not be shown here.

The MEM analysis has been performed with the continuum free spectral
function $\omega^2$ as default model, using the euclidean correlators
in a time window starting at  $\tau=2$, and cutting off the energy
integral in (\ref{eq:spectral}) at $a_t\omega_{\text{max}}=6$.

\section{Results}

Figure \ref{fig:polyak} shows the average Polyakov loop $\bra L\ket$
as a function of $1/N_t$.  We find that with our small volume and large
lattice spacing it is not possible to determine the critical
temperature, since no region exhibits a particularly rapid change in
this quantity.  In agreement with this, the Polyakov loop
susceptibility has no discernable peak in this region.  Larger
lattices will be required to determine $T_c$.  In the absence of any
such determination we take $T_c$ to be in the region where $\bra
L\ket$ begins to assume a value significantly above zero.
\begin{floatingfigure}[r]
\includegraphics*[width=0.5\textwidth]{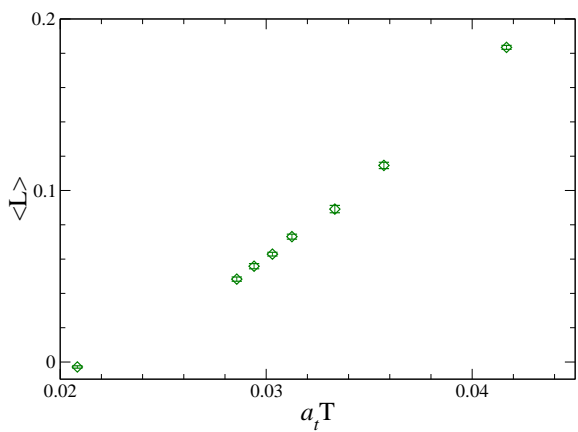}
\caption{Average Polyakov loop as a function of temperature.}
\label{fig:polyak}
\end{floatingfigure}

\begin{figure}
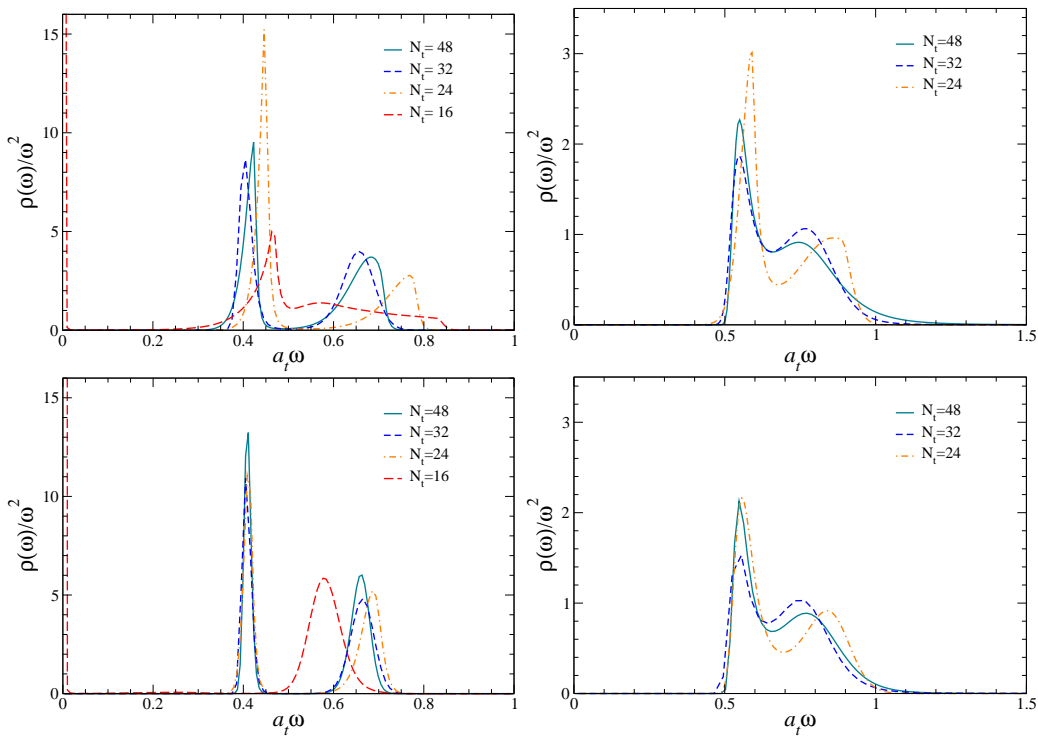

\includegraphics*[width=0.45\textwidth]{rho_p5_m01.eps}
\includegraphics*[width=0.45\textwidth]{rho_p5_m02.eps}\\
\includegraphics*[width=0.45\textwidth]{rho_vi_m01.eps}
\includegraphics*[width=0.45\textwidth]{rho_vi_m02.eps}
\caption{Pseudoscalar ($\eta_c$, top) and vector ($J/\psi$, bottom)
  spectral function for different temperatures and bare quark
  mass $a_tm_0=0.1$ (left) and 0.2 (right).}
\label{fig:spf}
\end{figure}

We have computed correlators for bare quark masses $a_tm_0=0.1,0.2$,
where 0.1 is close to but slightly lighter than the physical charm
quark mass.
Figure \ref{fig:spf} shows the spectral
function obtained from MEM for the pseudoscalar ($\eta_c$) and vector
($J/\psi$) channels.  The position of the main peak for $N_t=48$
agrees with the mass obtained for the respective particles on the same
lattices, using a variational basis of smeared operators
\cite{Juge:2005xx}.  The second peak cannot be identified
with the first radially excited state, which on these lattices is
found to be $a_tm'=0.52$ \cite{Juge:2005xx}, in agreement with the PDG
value for $\psi'$. It is most likely a combination
of lattice artefacts and contributions from excited states 2S, 3S etc,
which cannot be resolved by the present data.

\begin{figure}
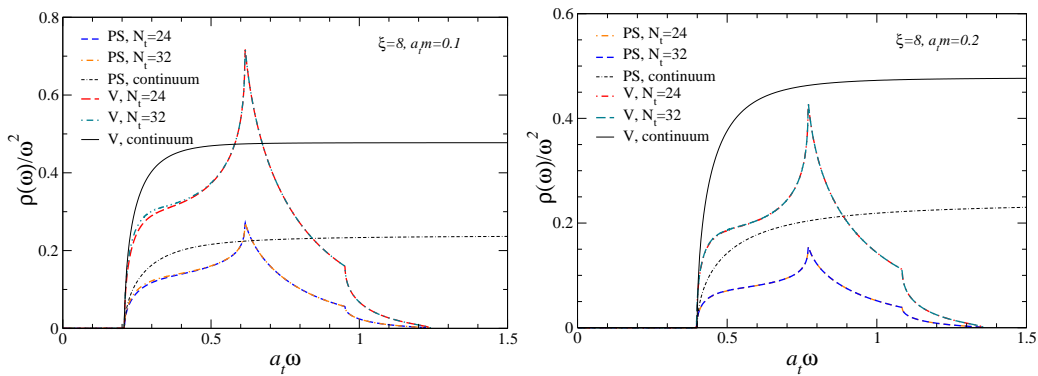

\includegraphics*[width=0.45\textwidth]{plot_TCD_m0.1.eps}
\includegraphics*[width=0.45\textwidth]{plot_TCD_m0.2.eps}
\caption{Free-fermion spectral functions in the pseudoscalar
  and vector channels, for bare quark
  mass $a_tm_0=0.1$ (left) and 0.2 (right).}
\label{fig:free}
\end{figure}

The issue of lattice artefacts can be addressed by comparing with the
free lattice spectral function, which is shown in fig.~\ref{fig:free},
together with the continuum free spectral functions.  The most
striking feature is the cusp at $a_t\omega\sim0.6 (0.77)$ for
$a_tm_0=0.1 (0.2)$, which coincides with the second peak in our data.
It is thus not possible to attribute any physical significance to the
second peak in our data.  The lattice spectral functions also
undershoot the continuum curve at an early stage, indicating that a
finer lattice is highly desirable.  Due to the lattice cutoff, the
free spectral functions go to zero for $a_t\omega\gtrsim1.25$, and it
would thus be sensible to also cut off the integral in
(\ref{eq:spectral}) near this point.

Taken at face value, the results shown appear to indicate that the 1S
states remain unchanged, or may even (in the case of $\eta_c$) become
more strongly bound, at temperatures 1--1.5$T_c$.  This would not be
entirely surprising in light of recent calculations of heavy quark
internal energies \cite{Kaczmarek:2005gi}.  However, it will be
necessary to investigate to what extent these results remain unchanged
with increased statistics.  We find some dependence on the range of
time separations used in the analysis; this may be statistics-related.

At $N_t=16$ the signal has changed substantially: the second peak has
disappeared, while the primary peak is substantially weakened and
shifted to higher energies, indicating a melting of the charmonium 1S
states.  There is also a non-zero signal at low $\omega$, but whether this
reults in a non-zero conductivity requires further investigation.

\section{Discussion and outlook}

The results presented here appear to confirm the picture emerging from
quenched simulations, that $J/\psi$ and $\eta_c$ survive in the medium
up to $1.5T_c$ or higher, while excited states begin to melt away at lower
temperatures.  There are even indications that $\eta_c$ might be more
strongly bound at intermediate temperatures, which, if true, would be
in accordance with recent results for the internal quark--antiquark
internal energy \cite{Kaczmarek:2005gi}.  No such changes are seen for
$J/\psi$, spoiling this agreement.  There are, however, a number of
issues that must be addressed before any conclusions can be drawn with
confidence.

An important source of systematic uncertainty is that the anisotropy
has not been completely tuned: the quark anisotropy is significantly
larger than the gluon anisotropy.  This can lead to the quarks
``feeling'' a higher temperature than the gauge fields.  It will be
important to repeat these calculations with a fully tuned parameter
set.

The lattice volume used in this initial study is quite small, at only
1.6 fm$^3$, so finite volume effects may have a substantial impact on
the results.  Future simulations will be carried out on a larger
volume.  It is also important to increase the statistics, in
particular if we wish to study the effects on higher excited states,
which will only be clearly resolved with higher statistics.  This is
also necessary if the effects of the zero-temperature decay width
(negligible for $J/\psi$), thermal width and effects of finite
statistics are to be resolved.

As the free spectral functions in fig.~\ref{fig:free} indicate, our
coarse spatial lattice means that lattice artefacts are substantial
even at relatively low energies.  Although this problem may be
ameliorated somewhat by using the free lattice spectral functions as
part of the prior knowledge, it will ultimately be necessary to repeat
the calculation at smaller lattice spacing.  This will however require
a new nonperturbative tuning process.

These results have been obtained using the continuum free spectral
function as default model.  We are planning to repeat our analysis
using other default models, in particular the free lattice spectral
function, to provide a check on the systematics of the MEM analysis.

%To do: compare correlators with reconstructed corrs from
%$\rho(\omega)$ at $N_t=48$ \cite{Datta:2003ww}.

In the future, we plan to extend this study to light vector meson
correlators at zero and non-zero momentum, which will yield
information about dilepton production rates in the plasma.  Lattice
artefacts are expected to be less severe at smaller quark masses, so
this may be feasible even on current lattices.  We also intend to
study Landau damping by investigating the behaviour of light- and
heavy-quark spectral functions at non-zero momentum below the
lightcone.

\section*{Acknowledgments}

This work has been supported by the IRCSET Embark Initiative award
SC/03/393Y, SFI grant 04/BRG/P0275 and the IITAC PRTLI initiative.  We
wish to thank Jimmy Juge and Simon Hands for stimulating and fruitful
discussions.

\bibliography{lattice,trinlat,hot}
\end{document}